\documentclass[epj]{svjour}
\usepackage{graphics,epsfig,amsfonts,amssymb,amsmath,subfigure,psfrag}
\def\Journal#1#2#3#4{{#1} {\bf #2} (#4) #3}

\def\PRD{{Phys. Rev.}    {\bf D}}

\begin{document}
\title{An Updated Description of Heavy-Hadron Interactions in {\sc Geant}-4}
%\subtitle{Do you have a subtitle?\\ If so, write it here}
\author{R. Mackeprang\inst{1} \and D.~A. Milstead\inst{2}% etc
% \thanks is optional - remove next line if not needed
%%%%\thanks{\emph{Present address:} Insert the address here if needed}%
}                     % Do not remove
%
%%%\offprints{}          % Insert a name or remove this line
%
\institute{CERN, Geneva, Switzerland \and Fysikum, Stockholms Universitet, Stockholm, Sweden.}
%
%%%\date{Received: date / Revised version: date}
% The correct dates will be entered by Springer
%
\abstract{
Exotic stable massive particles (SMP) are proposed in a number of
scenarios of physics beyond the Standard Model.  It is important that
LHC experiments are able both to detect and extract the quantum
numbers of any SMP with masses around the TeV scale. To do this, an
understanding of the interactions of SMPs in matter is required. In
this paper a Regge-based model of $R$-hadron scattering is extended
and implemented in {\sc Geant-4}. In addition, the implications of
$R$-hadron scattering for collider searches are discussed.
\PACS{
      {24.10.Lx}{Monte Carlo simulations}   \and
      {14.80.Ly}{Supersymmetric partners of known particles.}
     } % end of PACS codes
} %end of abstract
\maketitle

\section{Introduction}
The observation of an exotic stable\footnote{The term stable is taken
  to mean that the particle will not decay during its traversal of a
  detector.} massive particle (SMP) with electric, magnetic or
colour charge (or combinations thereof) would be of fundamental
significance. Searches are therefore routinely made for SMPs bound in
matter, and for SMPs at cosmic ray and collider
experiments~\cite{Amsler:2008zzb,Drees:1990yw,Perl:2001xi,Fairbairn:2006gg,Raklev:2009mg}.
Furthermore, SMPs with masses at the TeV scale are anticipated in a
number of beyond-the-Standard-Model scenarios, such as supersymmetry
and universal extra dimensions~\cite{Fairbairn:2006gg,Raklev:2009mg}. SMP searches
will thus be a key component of the early data exploitation programme of
the LHC experiments~\cite{Aad:2009wy,CMS-SMP,Pinfold:1999sp}. To facilitate
such searches, models are needed of the interactions of SMP as they
pass through matter. Although robust phenomenological approaches are
available to predict the scattering of electromagnetically charged
SMPs, it is not clear how the interactions of hadronic SMPs (hereafter
termed $R$-hadrons\footnote{The term $R$-hadron has its origin in the
  $R$-parity quantum number in supersymmetric theories and the work in
  this paper will be presented in the context of supersymmetric
  particles. However, the results given here are generally applicable to stable
  heavy exotic hadrons.}) should be treated. This is due to uncertainties
associated with the strong scattering processes and the hadronic mass
hierarchy.  In this work a model of $R$-hadron scattering~\cite{deBoer:2007ii} which is based
 on triple Regge phenomenology~\cite{Collins:1977jy} is extended and implemented in {\sc Geant-4}~\cite{Agostinelli:2002hh}. Although the main purpose of
 this work is to aid searches at colliders, it is also envisaged that it will be useful for
cosmic-ray searches~\cite{Anchordoqui:2007pn}.

Hadronic interactions of $R$-hadrons at colliders can give rise to energy loss and a range of
striking signatures. Scattering processes can thus potentially enhance
the discovery capability of an experiment as well as challenging it's
event-reconstruction algorithms~\cite{Fairbairn:2006gg}. For example, due to
scattering effects, a SMP can possess different values of electric
charge as it traverses a detector. Furthermore, hadronic interactions
affect the rate at which SMPs lose energy and thus the potential of
experiments to discover SMPs via the late decays of stopped
SMPs~\cite{Arvanitaki:2005nq,Abazov:2007ht}.

To aid the development of search strategies, it is thus important that
experiments have access to Monte Carlo models which span as fully as
possible the range of conceivable signatures which could be associated
with $R$-hadron production. At present, one model is available within
{\sc Geant-4}~\cite{Mackeprang:2006gx}. This model, hereafter termed
the \emph{generic model}, is based on a black-disk
approximation~\cite{Kraan:2004tz} and employs pragmatic assumptions
regarding possible scattering processes and stable $R$-hadron
states. The work presented in this paper concerns an approach, hereafter termed the
\emph{triple Regge model}, which uses the triple Regge formalism to predict the
scattering of $R$-hadrons formed from stop-like ($\tilde{t}$) and sbottom-like ($\tilde{b}$) exotic colour-triplet squarks ($\tilde{q}$)  and electrically neutral gluino-like ($\tilde{g}$) colour-octet particles~\cite{deBoer:2007ii}.
Furthermore, the model uses well motivated assumptions regarding
$R$-hadron mass hierarchies which differ from those employed by the
generic model.

This paper is organised as follows. First, we discuss mass hierarchies
of $R$-hadrons. This is followed by a description of the triple Regge model
and its implementation in {\sc Geant-4}. Finally, model calculations
of the energy loss and flavour composition for $R$-hadrons passing through
iron are shown and discussed.

\section{Exotic heavy hadrons and their masses}\label{sec:massh}
Fermionic mesons ($\tilde{q}\bar{q}'$,$\tilde{g}q\bar{q}'$)
and bosonic baryons \\
($\tilde{q}q'q''$,$\tilde{g}qq'q''$) would be expected to be formed from
stable squark and gluino states, where $q,q'$ and $q''$ denote Standard Model quarks~\cite{Farrar:1978xj}\footnote{Where no ambiguity is likely to arise, a reference to a particle is also taken to refer to its antiparticle in this paper.}. In this work, with the exception of the lightest gluino baryonic state (described below), only $R$-hadrons comprising light valence quarks of type $u$ and $d$ are considered.

Calculations of $R$-hadron mass spectra have been made using a variety
of approaches such as the bag model~\cite{Chanowitz:1983ci,Farrar:1984gk,Buccella:1985cs} and
lattice QCD~\cite{Foster:1998wu}. These predictions, together with
with measurements of heavy hadron masses, make it possible to
approximately determine those features of the mass hierarchies which
are most relevant for the modelling of $R$-hadron scattering in
matter. Of particular interest are the masses of the lowest-lying
states, to which higher mass particles would be expected to decay
before interacting hadronically, and the mass splitting between
the lightest meson and baryon states.

Given the observed spin and flavour independence of interactions of
heavy quarks with light quarks~\cite{Isgur:1989vq}, the lightest squark-based meson
and baryon states can be inferred~\cite{Gates:1999ei} from the
measured mass spectra of charm and bottom hadrons. Thus, using the
mass spectra of the observed charm and bottom mesons as a guide, charged and
neutral squark-based mesons would be approximately mass degenerate, and
the lightest baryon state would be the scalar singlet $\tilde{q}ud$. This
would be the exotic equivalent of the
$\Lambda_c$,$\Lambda_b$ baryons. This picture is supported by
calculations of hyperfine splittings of a few hundred MeV
between the $\tilde{q}ud$ state and the vector states
$\tilde{q}uu$,$\tilde{q}dd$ and
$\tilde{q}ud$~\cite{Kraan:2004tz,Gates:1999ei}. Furthermore, the mass
difference between the stable baryon and mesonic states would be far
less than 1 GeV and the decay of a baryon into a meson and proton
would be kinematically forbidden~\cite{Kraan:2004tz,Gates:1999ei}.

The basic properties of gluino $R$-hadrons may not be similarly
inferred from Standard Model hadronic mass spectra. A greater
reliance on phenomenological approaches is therefore required to identify the
stable states. Possible hadrons include gluino $R$-mesons
($\tilde{g}q\bar{q}^0$
,$\tilde{g}u\bar{d}^+,\tilde{g}d\bar{u}^-$)\footnote{Here $q$$\bar{q}$ represents a mixture
  of $u\bar{u}$ and $d\bar{d}$-states.} and gluino-gluon states
($\tilde{g}g$). Various calculations, such as in Ref.~\cite{Chanowitz:1983ci}
(bag model), Ref.~\cite{Foster:1998wu} (lattice QCD) and
Ref.~\cite{Kraan:2004tz} (Potential model) estimate the masses of the gluino $R$-mesons to be
the gluino mass plus 0.2 - 0.7 GeV.  The mass differences between the
lowest lying states of each of these hadrons are less than the pion
mass. Calculations of the lowest lying baryonic states using the bag
model~\cite{Farrar:1984gk,Buccella:1985cs} predict that the lightest
state is the flavour singlet $\tilde{g}uds^0$, and that the "nucleon"
$R$-baryons ($\tilde{g}udd^0$, $\tilde{g}uud^+$) are 0.2 - 0.3 GeV
heavier. The consequence of this is that even though these heavier
states would be stable to strong decays, they would decay weakly into
the singlet state over a time scale of $\sim 10^{-10}$ s. The mass of
the $\tilde{g}uds^0$ state is typically in the range of the gluino
mass plus 0.2 - 0.7 GeV~\cite{Chanowitz:1983ci,Buccella:1985cs}. It is
thus reasonable to assume that all gluino $R$-hadrons would be
approximately degenerate in mass. Specifically none of the models allow for a
$R$-meson / $R$-baryon mass splitting larger than the proton mass
thus disallowing decays from baryonic to mesonic states.

\begin{table}[h!]
{\begin{tabular}{|c|c|c|}
  \hline
  % after \\: \hline or \cline{col1-col2} \cline{col3-col4} ...
  Heavy Parton & States & Mass (GeV)\\
  \hline
  Squark & ${\tilde{q}\bar{u}},{\tilde{q}\bar{d}}$ &   $m_{\tilde{q}}+0.3$ \\
   &  $\tilde{q}{ud}$  & $m_{\tilde{q}}+0.7$  \\
  \hline
  Gluino & ${\tilde{g}q\bar{q}}$,${\tilde{g}u\bar{d},\tilde{g}d\bar{u}}$, ${\tilde{g}g}$ & $m_{\tilde{g}}+0.7$ \\
   & ${\tilde{g}uds}$  & $m_{\tilde{g}}+0.7$ \\
  \hline
\end{tabular}
\caption{Assumed stable hadrons formed from squarks ($\tilde{q}$) and gluinos ($\tilde{g}$), together with mass estimates
used in this work. The neutral gluino states containing a mixture $u\bar{u}$ and $d\bar{d}$ pairs
are generically denoted by ${\tilde{g}q\bar{q}}$.
%To estimate uncertainties, two values of gluino baryon mass are considered.
}\label{tab:tab1}}\end{table}

For the work presented here, we use the squark and gluino $R$-hadron
mass hierarchy assumptions given in Table~\ref{tab:tab1}. In addition, to avoid
attempting calculations of poorly understood cascade decays, we use
simplified mass spectra which comprise only those $R$-hadrons deemed
to be stable. The underlying assumption is that any higher states would decay to
the stable states before having an opportunity to interact.

%A further simplifying assumption used is that, with the exception
% of the lightest gluino
%$R$-baryon states, $R$-hadrons containing only light $u$, $d$ quarks
%were considered.
% The states assumed to be stable are summarised in
% Tab.~\ref{tab:tab1}.  In addition, as discussed later in
% Section.~\ref{sec:res}, the value of the lighest gluino baryon mass
% is increased by 400 MeV in order to estimate theoretical
% uncertainties associated with the assumed mass hierarchies.

Partly as a consequence of the assumed stable states, $R$-hadron event
topologies can be anticipated which would be challenging for one of
the most common collider search methods. The efficiencies of searches for slow
muon-like objects~\cite{Aad:2009wy,Abe:1989es,Abe:1992vr,Acosta:2002ju,Abazov:2008qu,Aaltonen:2009ke} would
be affected by the $R$-baryon production processes which take place as
$R$-hadron pass through matter. As discussed
further in Section~\ref{sec:scatt}, such processes imply that a sbottom or gluino
$R$-hadron, irrespective of its state as it entered a calorimeter, is
likely to leave as an uncharged object and therefore escape detection
in an outer muon detector.

Although well motivated mass-hierarchy assumptions were selected for
this work, in the absence of experimental data, alternative scenarios
should also be considered. The generic $R$-hadron scattering model
makes more pragmatic assumptions regarding the mass hierarchies. This
leads to the bulk of sbottom and gluino $R$-hadrons being charged
after they leave a typical calorimeter~\cite{Aad:2009wy}. Thus, taken
together, the two models will aid in the development of a more
complete set of collider searches for $R$-hadrons.

\section{Scattering of $R$-hadrons in Matter}\label{sec:scatt}
Although the fine details of $R$-hadron scattering are difficult to
model, first-principle arguments can be used to build up a qualitative
picture of the processes through which $R$-hadrons suffer energy loss
and undergo charge and baryon number exchange.

The probability of an interaction between the heavy parton and a quark
in the target nucleon is low since the cross section varies with the
inverse square of the parton mass according to perturbative
QCD~\cite{Kraan:2004tz}. One can thus use a picture of scattering of a
stable non-interacting heavy parton accompanied by a coloured hadronic
cloud of light constituents which take part in the scattering. Energy
losses of a $R$-hadron in an interaction with a stationary nucleon
will thus be determined by the kinetic energy of the light quark
system in a $R$-hadron. At the LHC, for $R$-hadrons of masses above
several hundred GeV the light quarks system's kinetic energy is typically of
order GeV and small energy losses are thus anticipated. Another
important feature of $R$-hadron scattering is that processes in which
$R$-baryons are converted into $R$-mesons are suppressed due to
kinematics and the absence of pions in
material~\cite{Kraan:2004tz}. Thus, any mesons which convert to
baryons will likely stay in this state during their passage through
material. A number of models have been
proposed~\cite{deBoer:2007ii,Mackeprang:2006gx,Kraan:2004tz,Baer,mafi}
which are based on the above principles. In this section, brief
summaries are given of the salient features of the
triple Regge~\cite{deBoer:2007ii} and generic
models~\cite{Mackeprang:2006gx,Kraan:2004tz}. This is followed by a
description of the implementation of the triple Regge model in {\sc Geant-4}.

\subsection{Triple Regge model}\label{sec:Regge}
Since the central picture is one of a low-energy light-quark system
interacting with a stationary nucleon, $R$-hadron scattering can be
treated with the phenomenology used to describe low-energy
hadron-hadron scattering data~\cite{deBoer:2007ii,Baer,mafi}, as is
done in the triple Regge approach~\cite{deBoer:2007ii}.  The triple Regge model was
originally developed to describe the scattering of exotic hadrons containing heavy colour-triplet objects. Here,
though, it has been extended to also treat gluino $R$-hadrons. This
model assumes the stable states described in Section~\ref{sec:massh}.

\begin{figure}{\epsfig{file=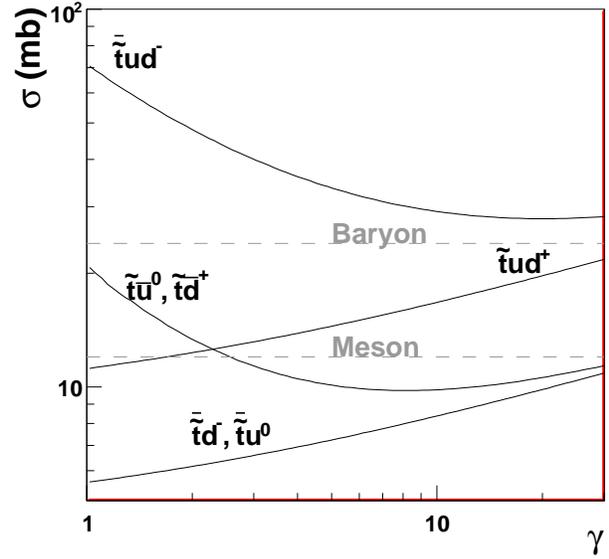,width=0.45\textwidth}\hfill
\epsfig{file=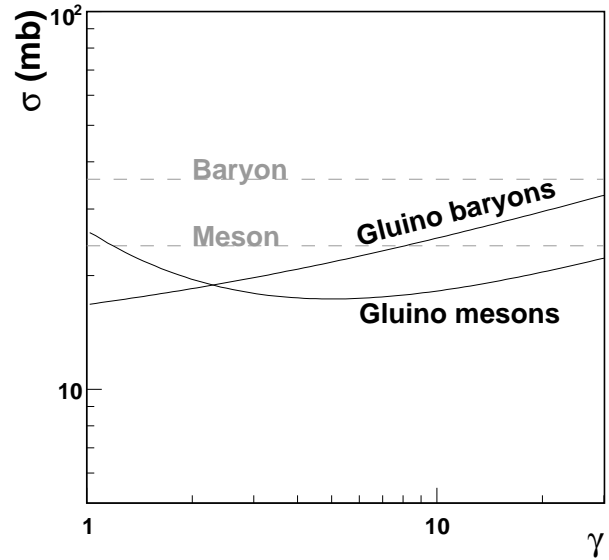,width=0.45\textwidth}
\caption{Cross sections for the interaction of stop-based and gluino-based $R$-hadrons off a stationary nucleon within a nucleus assumed to contain an equal number of protons and neutrons. The predictions are shown for the triple Regge model (solid lines) and the generic model (dashed lines).}
\label{fig_cx}}\end{figure}

Using parameters fitted to low-energy hadron-hadron data,
the triple Regge model makes predictions for $R$-hadron scattering cross sections,
together with energy-loss calculations based on the triple Regge
formalism.  Figure~\ref{fig_cx} (left) shows the the model predictions
of the scattering cross sections of a squark-based $R$-hadron off a
stationary nucleon within a nucleus comprising equal numbers of
neutrons and protons. The cross section formulae are given in Section~\ref{sec:imp} in which the
implementation of the model in {\sc Geant-4} is described. The cross section is shown for different types
of squark-based $R$-hadrons as a function of the Lorentz factor,
$\gamma$. As can be seen, there is a large cross section for
antibaryon ($\bar{\tilde{q}}\bar{u}\bar{d}$) interactions which is
due to a dominant annihilation process with a nucleon in the target.

It is also seen that, at lower values of $\gamma$ ($\gamma \lesssim
10$), the scattering cross section of squark-based $R$-hadrons
containing a light valence antiquark
($\tilde{q}\bar{u},\tilde{q}\bar{d}$) is larger than for
antisquark-based $R$-mesons. This arises from the presence of Reggeon-exchange
 processes which are only permitted for $R$-hadrons containing
a light antiquark. Owing to the presence of an additional light
quark the scattering cross section for $R$-baryons
(${\tilde{q}}{u}{d}$) is twice as large as for the mesons with light
quarks.

\begin{figure}{
\epsfig{width=7cm,file=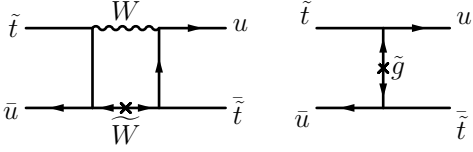}
\caption{Feynman diagrams representing oscillations of neutral stop-based $R$-mesons.}
\label{figfeyn}}\end{figure}

Upon an interaction the probability that a $R$-meson becomes a baryon
is around 10\%. Once a $R$-hadron becomes a baryon it stays in this
state. Another process which must be taken into account is the
oscillation of a neutral mesonic squark-based $R$-hadron into its
antiparticle~\cite{Gates:1999ei,Sarid:1999zx}. Feynman diagrams of possible processes in which oscillations can occur
are shown in Figure~\ref{figfeyn}. Tree level gaugino (gluino and neutralino) exchange and one-loop charged current-chargino box
diagrams are shown. Since the conversion rate would be model dependent we
allow two possibilities here: zero mixing, in which no oscillations
take place and a maximal-mixing scenario in which there is a 50\%
probability that any neutral mesonic squark-based $R$-hadron which was
produced would automatically be converted to its anti-particle. These mixing scenarios correspond
to oscillation lengths which are infinite and zero, respectively.

In Figure~\ref{fig_cx} (right) the predicted cross sections for gluino
$R$-hadron species are shown. The baryon cross section is 50\% greater
than that for stop $R$-hadrons owing to the extra light quark in the
gluino $R$-hadron. Similarly, since the gluino meson contains a light
quark and antiquark, then both Regge and pomeron-exchange processes
are available. The scattering cross section for gluino mesons at a
fixed value of $\gamma$ is the sum of the cross sections for a stop
meson and anti-stop meson. Following Ref.~\cite{Kraan:2004tz}, a gluino-gluon state
is assumed to scatter in the same way as a gluino-based $R$-meson.  Again, a baryon number transfer probability
of 10\% is used.
%%A probability of $\frac{1}{3}$ is assigned to $R$-meson to $R$-meson processes involving
%%charge exchanges of 0, $\pm e$ or $\pm 2e$.

\subsection{Generic model}
The generic model is already available in {\sc
  Geant-4}~\cite{Mackeprang:2006gx} and is an update of earlier
work~\cite{Kraan:2004tz}. In view of the inherent uncertainties
associated with modelling $R$-hadron scattering a pragmatic approach
is assumed in which the scattering rate is estimated with a constant
geometric cross-section of 12~mb per light quark. All 2-to-2 and 2-to-3
processes are allowed if they are kinematically feasible and charge
conservation is respected. The proportion of 2-to-2 and 2-to-3
reactions is governed by phase space factors. The constant cross
sections for stop and gluino $R$-meson and $R$-baryon interactions are
superimposed on Figure~\ref{fig_cx} (left and right). This model
assumes $R$-hadron mass hierarchies which differ from those used in
the triple Regge model and predicts
that the majority of baryons which are produced in scattering
processes are electrically charged.

\subsection{Implementation of triple Regge model in {\sc Geant-4}}\label{sec:imp}
In the dynamical picture of $R$-hadron scattering used by {\sc Geant},
the light quark system is decoupled from the heavy spectator parton
before interacting with a nucleon according to {\sc Geant}'s
parameterised model for light hadrons.  In this way secondary
particles and so-called black tracks are generated. Following the
interaction, the light-quark system is recombined with the heavy
parton.

To implement the triple Regge model in {\sc Geant-4}, the software
architecture used for the generic model was adapted. The scattering
cross section was adapted from Ref.  \cite{deBoer:2007ii}. The pomeron
and Regge parts of the cross section per nucleon are (in mb):
\begin{eqnarray}
  \textrm{Reggeon} &:&  \sigma_R = \frac{3}{2} \exp\left[2.17+\frac{0.147}{\gamma}-0.961*\log{\gamma}\right]\\
  \textrm{Pomeron} &:&  \sigma_P = 4.14 + 1.50\sqrt{\gamma} - 0.0545\gamma + 0.000822\gamma^{3/2}
\end{eqnarray}
The scattering cross sections for squark and gluino $R$-hadrons\footnote{Gluino antibaryons
are not considered here for several reasons. The probability of the formation of an antibaryon  following the hadronisation of a gluino is predicted to be small
($\sim 1.5$\%)~\cite{Fairbairn:2006gg}. Furthermore, the probability of a gluino-based $R$-meson acquiring negative baryon number state in an
interaction in matter would also be expected to be strongly suppressed.} become:
\begin{eqnarray}
  \sigma(\tilde{q}\bar{q'}) &=& \sigma_P+\sigma_R\\
  \sigma(\tilde{q}ud) &=& 2\sigma_P\\
  \sigma(\tilde{g}q\bar{q'}) &=& \sigma_R+2\sigma_P\\
  \sigma(\tilde{g}uds)  &=& 3\sigma_P\\
  \sigma(\overline{\tilde{q}}q') &=& \sigma_P\\
  \sigma(\overline{\tilde{q}}\bar{u}\bar{d}) &=& 2(\sigma_P+\sigma_R)+\frac{30}{\sqrt{\gamma}}
\end{eqnarray}
The available baryon states were chosen as shown in table
\ref{tab:tab1} and the relative rates of the available processes
reweighted to match the prescription in Section~\ref{sec:Regge}. This
meant enforcing a 10\%{} probability of exotic meson $\rightarrow$
baryon conversion as well as disabling the phase space weights of the
generic model. The code is available for download \cite{webpage}.

Using single particle simulations, it was verified that the main
features of the triple Regge model were well reproduced in the {\sc
  Geant-4} implementation.  After the process reweighting, baryon
number changing processes formed 10\% of all meson interactions.  The
required charge exchange probabilities for meson-to-meson
transformations were guaranteed by the degeneracy of the $R$-mesons
and the availability of the same number and type of channels to every
meson in a multiplet. It was also confirmed that the {\sc Geant-4}
version reproduces the expected energy loss and charge exchange rates.

%\begin{itemize}
%\item Generic
%\item Regge
%\item Other ideas
%\item Resonances
%\end{itemize}

%\section{Prediction of Energy Loss and scattering}
%\subsection{Energy loss and multiple scattering}
%\subsubsection{Charge Exchange}

%\subsection{Implications on current limits}
%\begin{itemize}
%\item sbottom
%\item gluino new limit from the Tevatron
%\end{itemize]

\section{Results}\label{sec:res}
To study the expected effects of scattering on $R$-hadrons,
simulations were made of the passage through iron of $R$-hadrons of
300 GeV mass. The kinematic distributions of the $R$-hadrons were
given by the {\sc Pythia} generator~\cite{Sjostrand:2006za} which
simulated the direct pair production of $\tilde{t}\bar{\tilde{t}}$,
$\tilde{b}\bar{\tilde{b}}$ and $\tilde{g}{\tilde{g}}$ in proton-proton
collisions at 14 TeV centre-of-mass energy. The process through which
the partonic final state was hadronised and $R$-hadrons were formed
was modelled with customised {\sc Pythia}-routines~\cite{R-hadronweb}
based on the Lund string model~\cite{Andersson:1983ia}. A noteworthy
free parameter in this program is the probability of the formation a
$~\tilde{g}g$-state following the hadronisation of a gluino. This was
set to the default value of 0.1~\cite{R-hadronweb}. The influence of
the value of this free parameter on the results shown in this Section
is mentioned below.

For the results shown in this section, unless
otherwise stated, the allowed $R$-mesons and baryons were assigned the
masses prescribed in Table~\ref{tab:tab1}.

%\FIGURE[h]{\epsfig{width=7cm,file=E_loss_models_T.eps}
%\caption{Average energy loss per hadronic collision. Predictions of the generic and
%triple Regge model are shown. The effects of mixing are also shown.}
%\label{figecomp}}
\begin{figure}[h]{\epsfig{width=7cm,file=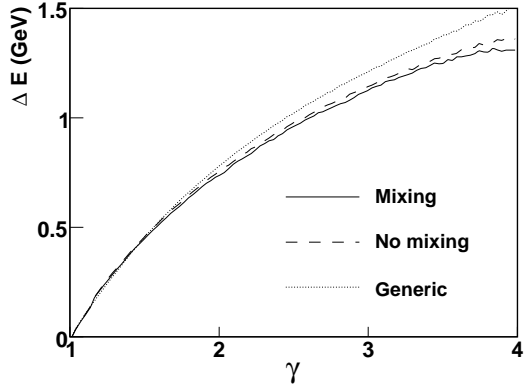}
\caption{Average energy loss per hadronic collision. Predictions of the generic and
triple Regge model are shown The influence of mixing is also shown.}
\label{figecomp}}\end{figure}

Figure~\ref{figecomp} shows the predicted average energy loss per
hadronic collision for squark-based $R$-mesons. There are only small
differences between the generic and triple Regge models. The energy losses of
squark and gluino-based $R$-baryons in both approaches (not shown) are
similar to the distributions shown in Figure~\ref{figecomp}.
%\FIGURE[h]{
%\epsfig{width=0.75\textwidth,file=eloss2mnew_label.eps}
%\caption{Total energy loss of different types of $R$-hadrons after passing
%through 2m of iron.  The influence of mixing is also shown.}
%\label{fig_eloss}}

\begin{figure}[h]
%\subfigure[Legend]{\epsfig{file=flavours_legend.eps,width=0.45\textwidth}}
\centering
\subfigure[\label{fig_2miron:eloss}]{\epsfig{file=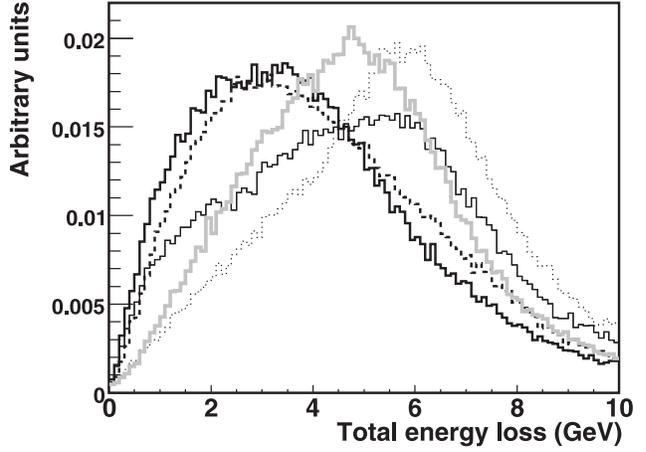,width=0.48\textwidth}}
\subfigure[\label{fig_2miron:nint}]{\epsfig{file=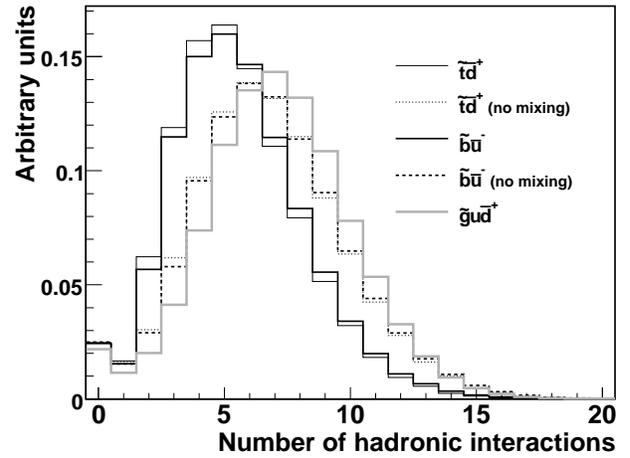,width=0.46\textwidth}}
\caption{Total energy loss (a) and number of hadronic interactions (b) for different types of $R$-hadrons after
propagating through 2m of iron.}
\label{fig_2miron}
\end{figure}

The total energy loss for $R$-hadrons passing through 2m of iron
 is shown in Figure~\ref{fig_2miron:eloss}. The length of 2m was chosen since it corresponds
 to the typical depth of a calorimeter at a collider~\cite{Wigmans:2000vf}.
 Gluino, stop and sbottom-based $R$-hadrons which start in the states
$\tilde{t}\bar{d}^+$,$\tilde{b}\bar{u}^-$, and $\tilde{g}u\bar{d}^+$,
respectively are shown.

 For a zero-mixing hypothesis, the stop-based
$R$-hadron suffers an average energy loss ($\sim 5.5$ GeV), which is
substantially greater than the energy losses for sbottom and
gluino-based $R$-hadrons.  This effect is due to the different
$R$-mesons converting into baryonic states. Of the different stable
baryon states, only the stop-based $R$-baryon is charged and therefore
able to lose energy through continuous ionisation ; conversions into baryon states are
studied in more detail below. As can be seen in Figure~\ref{fig_2miron:eloss}, the energy loss for the
gluino-based $R$-hadron is slightly larger than that for the sbottom
case owing to the typically greater number of hadronic interactions in
the gluino case, as seen in Figure~\ref{fig_2miron:nint}. It is also shown that mixing has
only a small effect on the energy loss and the number of hadronic interactions of the squark-based $R$-hadrons.

%\FIGURE[h]{
%\epsfig{width=9cm,file=nint2m.eps}
%\caption{Total number of hadronic interactions of different types of $R$-hadrons after passing
%through 2m of iron.  The influence of mixing is also shown.}
%\label{fig_nint}
%}

\begin{figure}[p]
%\subfigure[Legend]{\epsfig{file=flavours_legend.eps,width=0.45\textwidth}}
%%\hspace{-3.2cm}
\subfigure[Gluinos\label{fig_allflav1:g}]{\epsfig{file=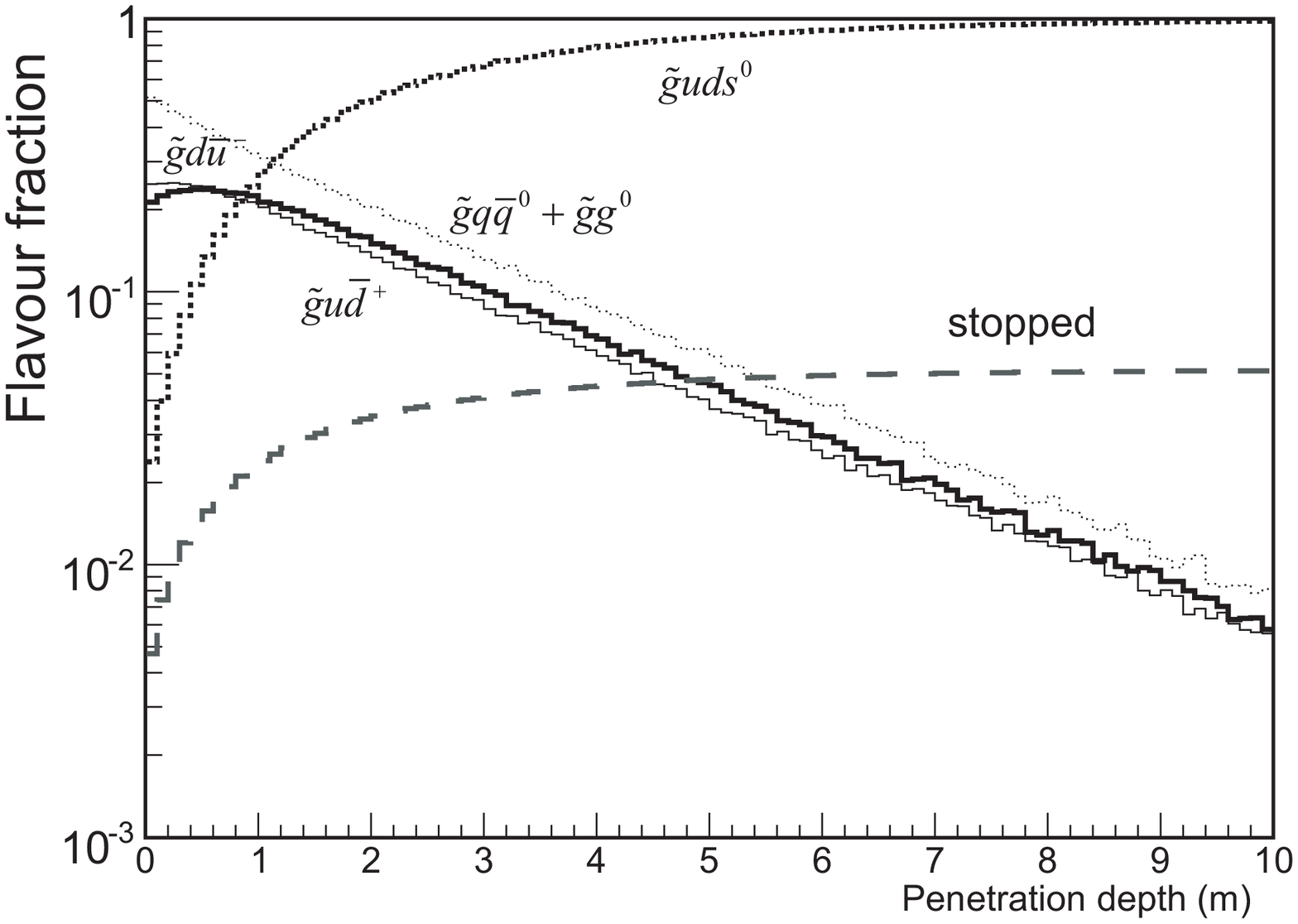,width=0.3\textwidth}} \\
%%\centering
\subfigure[Stops, no mixing\label{fig_allflav1:tnm}]{\epsfig{file=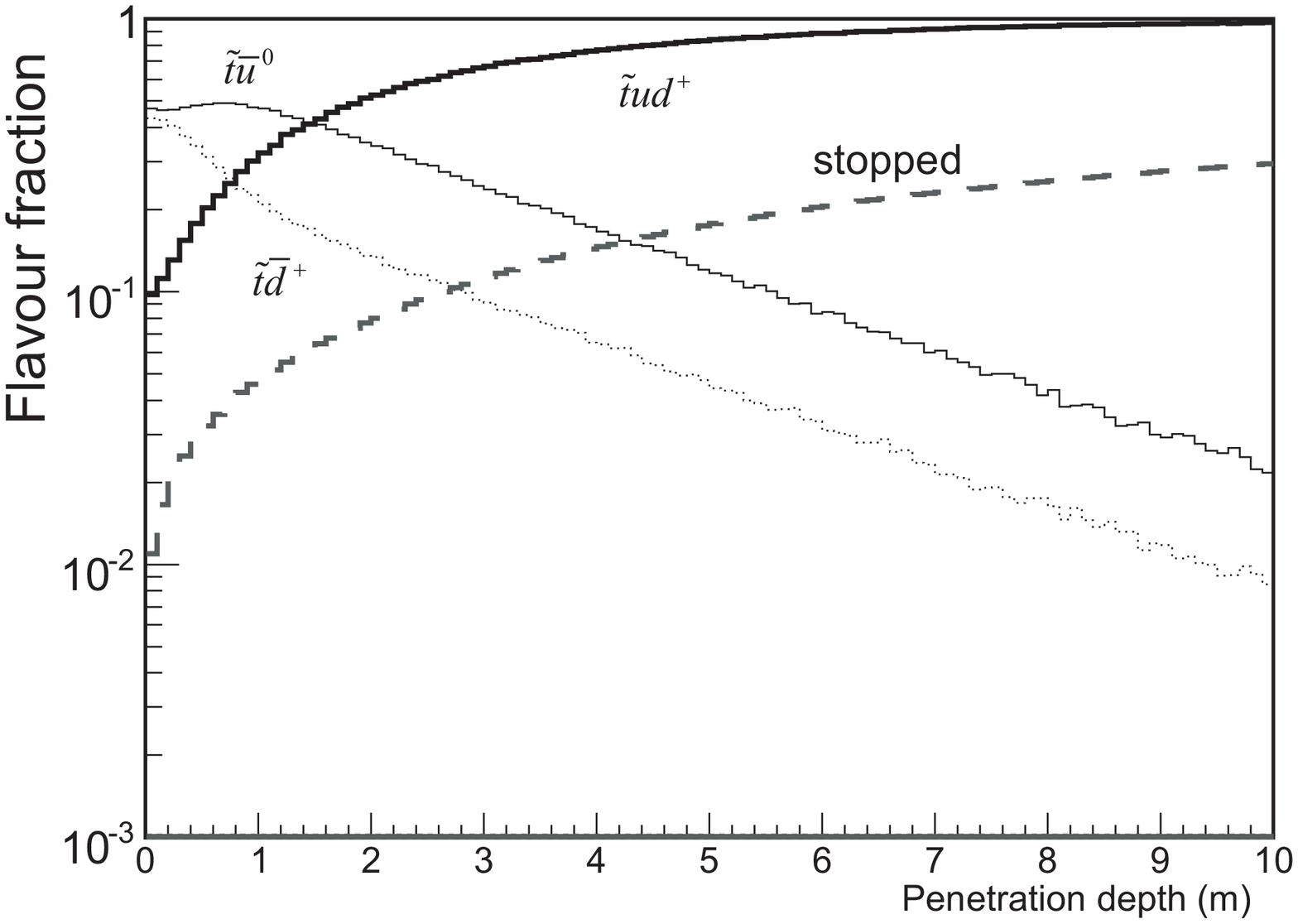,width=0.3\textwidth}}
\subfigure[Sbottoms, no mixing\label{fig_allflav1:bnm}]{\epsfig{file=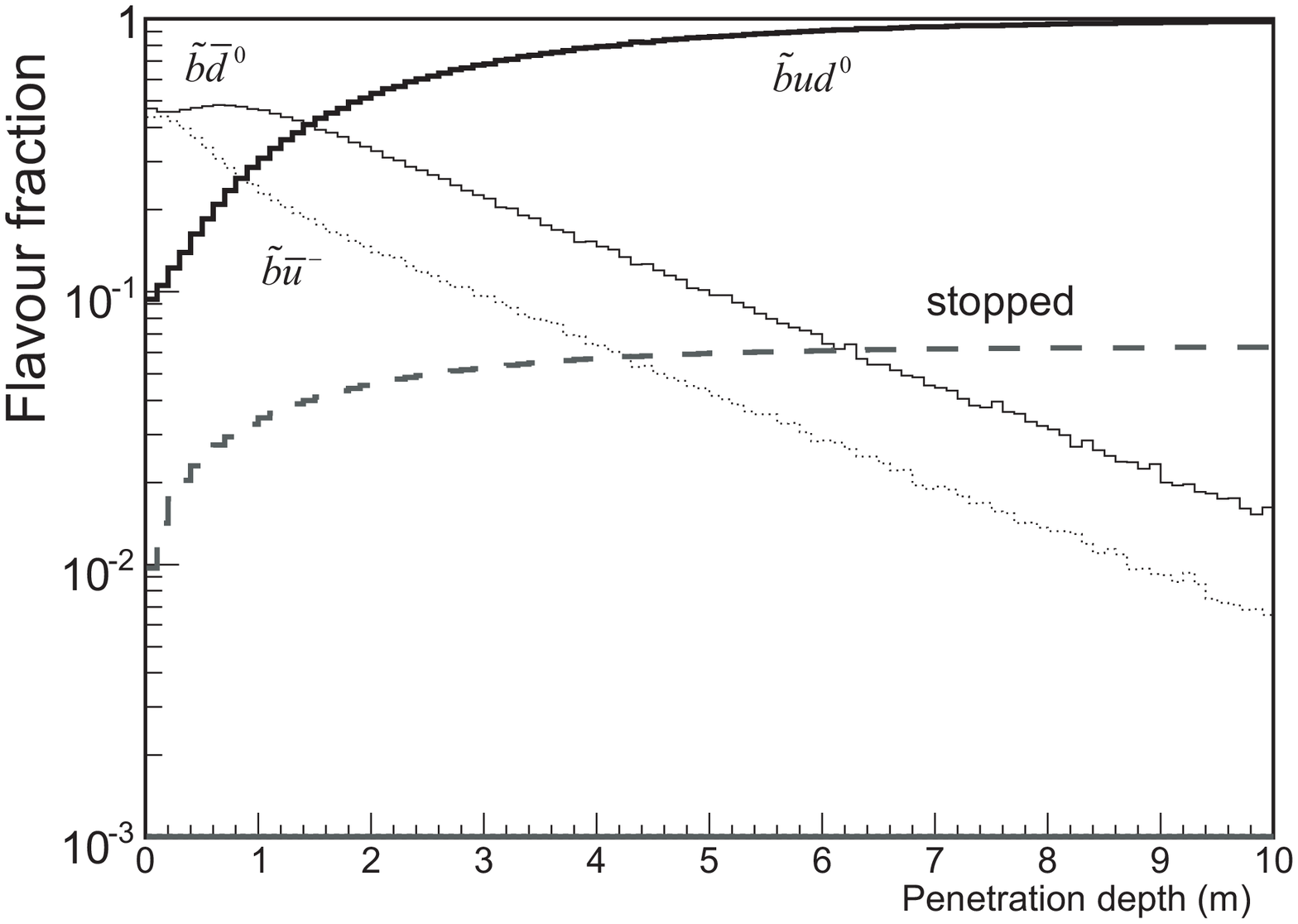,width=0.3\textwidth}}
\subfigure[Stops\label{fig_allflav1:t}]{\epsfig{file=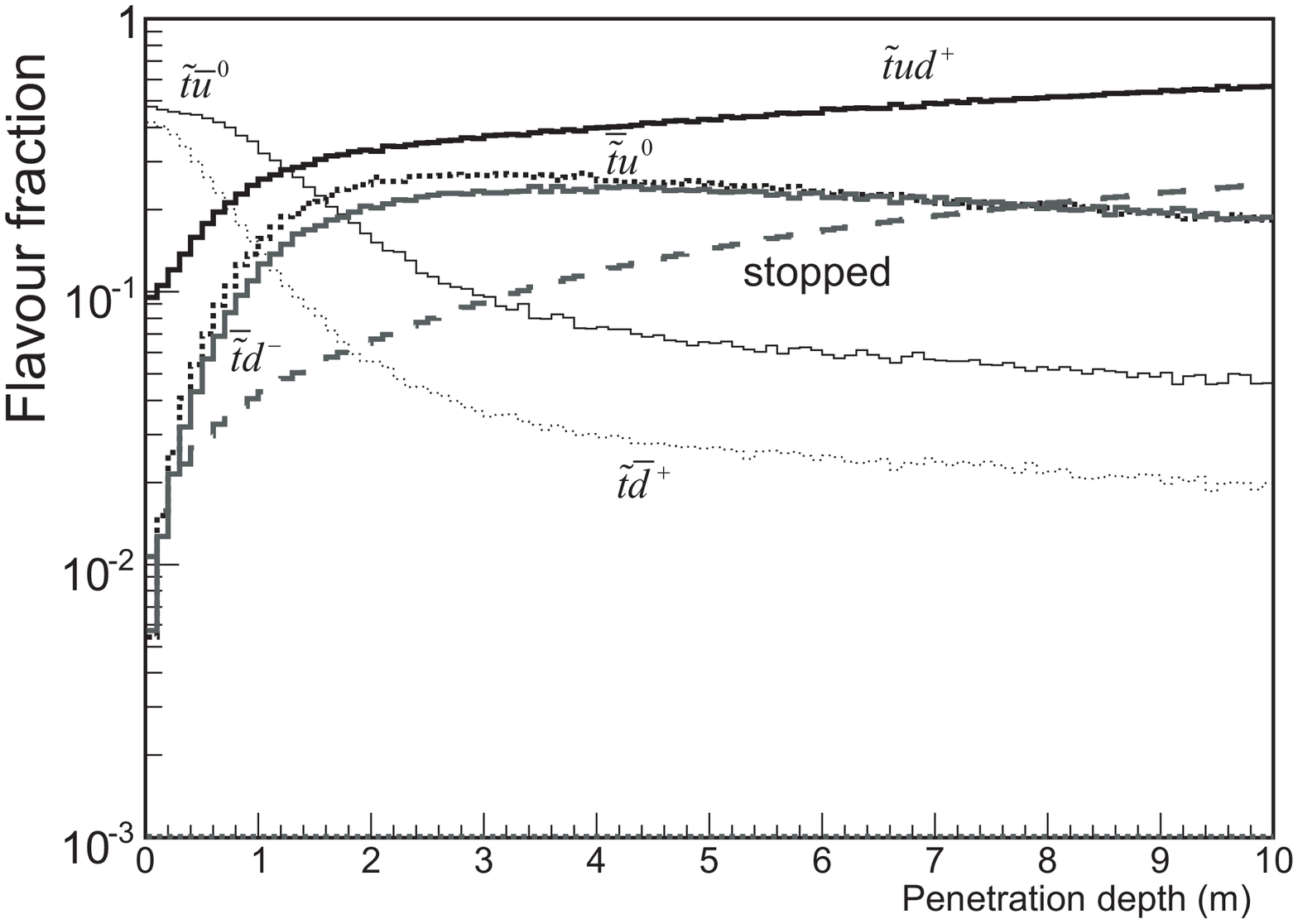,width=0.3\textwidth}}
\subfigure[Sbottoms\label{fig_allflav1:b}]{\epsfig{file=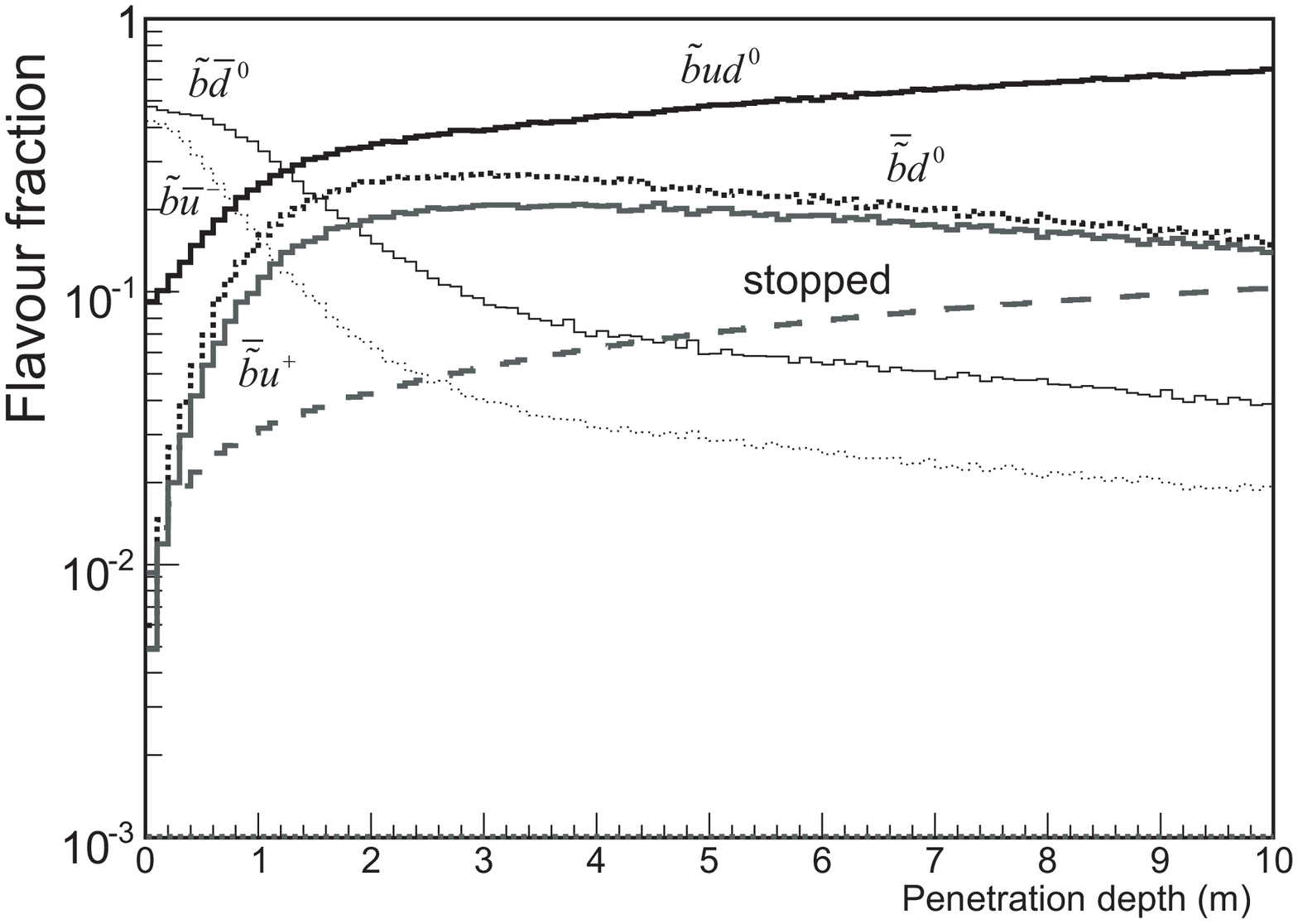,width=0.3\textwidth}}
\caption{Flavour composition of $R$-hadrons as a function of penetration depth in iron. Distributions are shown for
$R$-hadrons formed from gluinos, stops and sbottoms. The results of zero and maximal-mixing hypotheses are shown.}
\label{fig_allflav1}
\end{figure}

While $R$-hadron energy loss is a useful possible signature in a
collider search, the charge of the $R$-hadron after passing through
matter is of arguably greater importance to the development of search
strategies. As mentioned earlier, this is due to searches typically looking for
muon-like candidates emerging from thick calorimeter material. In Figure~\ref{fig_allflav1} the
proportions of different types of $R$-hadrons are shown as a function
of penetration depth in iron.

The spectra are shown for gluino, stop, and sbottom-based
$R$-hadrons. Prior to traversal, the proportions of different hadronic
states for a given heavy parton were calculated by the Lund string
model as mentioned above.
 As can be seen, the fact that $R$-hadrons are
most likely to be baryonic after traversing around 2m of iron, is
explicitly shown here. Of the remaining mesonic $R$-hadrons, around
one half would be electrically neutral. Thus, the majority of gluino
and sbottom-based $R$-hadrons would be neutral after a passage of
several metres in iron.

Varying the free parameter representing the probability of
 producing a $~\tilde{g}g$-state to the extreme values of $0$ and $1$ would clearly affect the relative ratios of
 different gluino-based $R$-hadron states formed in the primary interaction. However, the conclusion that a neutral baryon would be the dominant state after $\sim 2$m of iron would be unaffected since a $~\tilde{g}g$-state is assumed to scatter in the same way as a gluino-based $R$-meson as mentioned in section \ref{sec:Regge}.

The effects of maximal mixing, while tending
to slow down the meson-to-baryon conversion rate, similarly do not affect the conclusion that
squark-based $R$-mesons will dominantly be converted to baryons after travelling $\sim 2$m of iron. It is also implied in Figure~\ref{fig_allflav1} that $R$-hadrons can reverse the sign of their charge. For stop-based
hadrons this occurs through mixing, whereby a positively charged meson
can interact to become a neutral state which oscillates into its
antiparticle before interacting again to become a negatively charged
mesons. Furthermore, charge ``flipping'' can also occur for gluino-based
$R$-hadrons since positively and negatively charged mesons can both be
produced.  Also shown is the fraction of the $R$-hadrons which give up
all of their energy and are stopped in material.  After 2m of iron,
the fraction of stopped $R$-hadrons is typically 3-7\% depending on
the type of $R$-hadron and the mixing assumption.

\begin{figure}[p]
\centering
\subfigure[Antistops, no mixing]{\epsfig{file=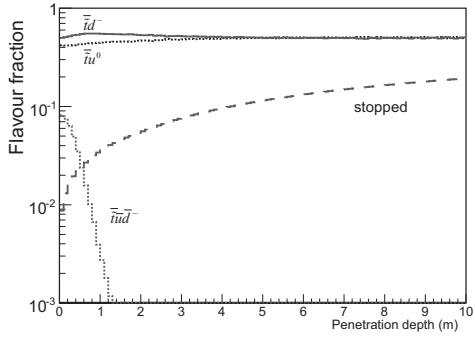,width=0.35\textwidth}}
\subfigure[Antisbottoms, no mixing]{\epsfig{file=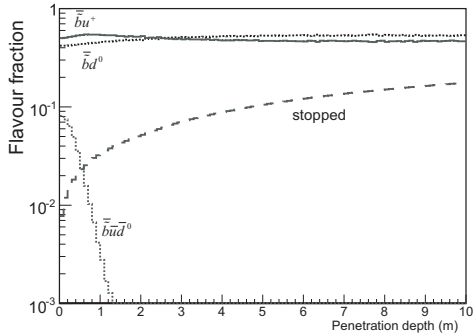,width=0.35\textwidth}}
\subfigure[Antistops]{\epsfig{file=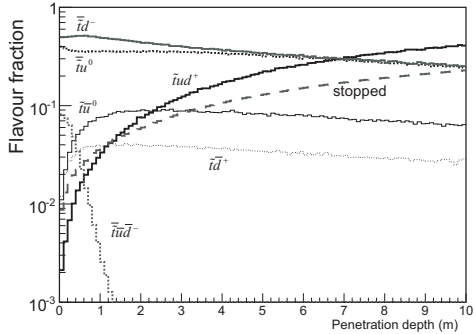,width=0.35\textwidth}}
\subfigure[Antisbottoms]{\epsfig{file=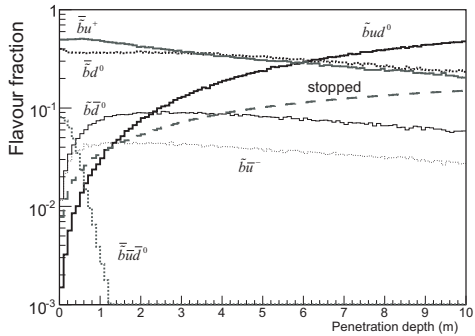,width=0.35\textwidth}}
\caption{Flavour composition of $R$-hadrons as a function of penetration depth in iron. Distributions are shown for
$R$-hadrons formed from antistops and antisbottoms. The results of zero and maximal-mixing hypotheses are shown.}
\label{fig_allflav2}
\end{figure}

Analogous flavour-composition distributions are shown in
Figure~\ref{fig_allflav2} for antistop and antisbottom-based
$R$-hadrons. As expected, the antibaryon states quickly die out owing
to the large annihilation cross section and, in the absence of mixing,
charged and neutral mesonic states are expected in roughly equal
proportions. The effects of mixing allow the formation of stop and
sbottom-based baryons and the same trends are thus observed as in
Figure~\ref{fig_allflav1}, albeit over longer traversal distances.

\section{Implications of $R$-hadron Scattering on Searches at Colliders}
Searches for stable $R$-hadrons at the Tevatron have exploited the expected penetrating muon-like behaviour of $R$-hadrons~\cite{Abe:1989es,Abe:1992vr,Acosta:2002ju,Abazov:2008qu,Aaltonen:2009ke}.
The most recent such search was made by the CDF experiment which reported
a lower mass limit for stable stop squarks of around 250
GeV~\cite{Aaltonen:2009ke}. It is, however, not trivial to use this result to estimate
mass limits on stable sbottom or gluino $R$-hadrons since, as discussed in this paper, such particles could
dominantly enter a muon chamber as electrically neutral objects.

There has so far been only one SMP-search~\cite{Abe:1992vr}  by a Tevatron experiment at which limits were estimated for heavy stable fermionic particles with colour octet charge\footnote{ The D$0$ experiment has looked for the late decays of long-lived gluino-based $R$-hadrons which would have stopped in it's calorimeters~\cite{Abazov:2007ht}. As for the non-decaying case, the obtained limits are sensitive to the assumptions made on the charge of $R$-hadrons in material.}. Under the assumption that the SMP would be electrically
charged as it propagates through the detector this work excludes stable gluinos of masses less than around 200
GeV. Also using this assumption, a recent SMP-search at CDF~\cite{Acosta:2002ju} has been interpreted in Ref.~\cite{Hewett:2004nw} as providing a lower mass limit of $\sim 300$ GeV.

If gluino $R$-hadrons do not behave as
muon-like objects then other channels need to be investigated to estimate limits.
CDF-measurements of events containing jets and missing
transverse energy~\cite{Acosta:2003tz} were used to estimate
a lower
mass limit on stable gluinos of around 170 GeV~\cite{Hewett:2004nw}. However, this limit
may only be relevant only for cases in which the $R$-hadrons are
uncharged during their complete traversal of the CDF detector since
the search applies a selection criteria which removes events
containing isolated high momentum tracks. Similarly, the recent D$0$
measurement~\cite{Abazov:2003gp} of the rate of events with the
monojet and missing transverse energy signature also applies selections
which may suppress $R$-hadron events; in this case via the rejection
of events containing calorimeter-muon candidates.

To make a rigorous estimate of gluino $R$-hadron mass limits requires dedicated detector simulations using several scattering models
to estimate the effects of the various event selections. However, in the absence of this it can nevertheless be argued
that the mass limits for stable gluinos may be well below 200 GeV. In a pessimistic scenario, lower mass limits on stable gluinos  would be $\sim 30$ GeV~\cite{Heister:2003hc,Abdallah:2002qi}, as obtained at LEP-2. At LEP searches typically assumed the production of a substantial rate of electrically charged $R$-hadrons and information from inner tracking devices was primarily used. Similarly, the mass limits on stable sbottom-like objects would be around 90 GeV~\cite{Heister:2003hc,Abreu:1998jy}.

Hadronic interactions and the properties of the lightest $R$-hadrons can thus play a crucial role in determining
detector signatures. Given the uncertainties on the properties of $R$-hadrons both at production and during their traversal of a detector it is therefore important that LHC experiments utilise fully the capabilities of their various subdetectors to look for a number of signatures.

%A further implication of hadronic scattering on collider searches concerns those searches
%which look for the late decay of stopped SMPs~\cite{Arvanitaki:2005nq}. Such work has already
% been undertaken by the D0 experiment~\cite{Abazov:2007ht} in the context of long-lived gluinos
% in a Split-Supersymmetry scenario. Since the fraction of stopped $R$-hadrons is dependent on assumptions
%  made in the scattering model, existing limits

\section{Summary}
An implementation and extension of model based on triple Regge phenomenology to squark
and gluino-based $R$-hadron scattering has been implemented in {\sc
  Geant-4}. The expected energy loss and the transformation of
$R$-hadrons as they propagate through material was studied. The model is complementary
to an existing {\sc Geant-4} approach for $R$-hadron scattering and will aid the development
of search strategies based on a range of different signatures associated with
 $R$-hadron production.

The implications of $R$-hadron scattering on the results of previous searches were also discussed. In the context of the model discussed in this work, it was shown that mass limits on different types of $R$-hadrons may be lower than previously thought.

The location of the {\sc Geant-4} code together with instructions for running it can be found in Ref.~\cite{webpage}.

\section*{Acknowledgements}

The authors wish to thank Nick Ellis, Beate Heinemann, Alexei Kaidalov and Terry Sloan for helpful comments and fruitful discussions. We are also very grateful to Glennys Farrar for useful discussions regarding calculations of $R$-hadron masses made with the bag model~\cite{Farrar:1984gk,Buccella:1985cs}.

David Milstead is a Royal Swedish Academy Research Fellow supported by a grant
from the Knut and Alice Wallenberg Foundation.

\end{document}